\documentclass[conference]{IEEEtran}
\usepackage{amsmath,amsfonts}
\usepackage{wrapfig}
\usepackage{graphicx}
\usepackage{setspace}
\usepackage{tabularx}
\usepackage{float}   
\usepackage{multicol}
\usepackage{indentfirst}
\usepackage{cite} 
\usepackage{caption}
\usepackage[final]{pdfpages}
\usepackage{tikz}
\usetikzlibrary{tikzmark}
\usepackage{footnote}
\usepackage{textcomp}
\usepackage{algorithm}
\usepackage{algorithmic}
\usepackage{multirow}
\usepackage{url}
\usepackage[hang,flushmargin]{footmisc}
\usepackage[super]{nth}
\usepackage{subcaption}
\usepackage{fancyhdr}
\usepackage{bm}
\usepackage{newfloat}
\usepackage{epsfig}

\usepackage{enumitem}
\usepackage{breqn}
\usepackage[export]{adjustbox}

\def\BibTeX{{\rm B\kern-.05em{\sc i\kern-.025em b}\kern-.08em
		T\kern-.1667em\lower.7ex\hbox{E}\kern-.125emX}}

\IEEEoverridecommandlockouts
\begin{document}
	\begin{spacing}{1}
		
		\title{A Configurable BNN ASIC using a Network of Programmable Threshold Logic Standard Cells
			\vspace{-15pt}}
		\author{\IEEEauthorblockN{Ankit Wagle\IEEEauthorrefmark{1}, Sunil Khatri\IEEEauthorrefmark{2}, Sarma Vrudhula\IEEEauthorrefmark{1}}
			\IEEEauthorblockA{\IEEEauthorrefmark{1} School of Computing, Informatics and Decision Systems Engineering, Arizona State University, Tempe AZ}
			\IEEEauthorblockA{\IEEEauthorrefmark{2} Dept. of Electrical and Computer Engineering, Texas A\&M University, College Station TX}\\[1em]
			\thanks{\IEEEauthorrefmark{1}The research was supported in part by NSF PFI award 1701241.}
			\vspace{-50pt}    
		}
		\maketitle
		
		\begin{abstract}
			This paper presents {\sc \textbf{Tulip}}, a new architecture for a binary neural network (BNN) that uses an optimal schedule for executing the operations of an arbitrary BNN. It was constructed with the goal of maximizing energy efficiency per classification. At the top-level, {\sc \textbf{Tulip}} consists of a collection of \textit{unique} processing elements ({\sc \textbf{Tulip}}-PEs) that are organized in a SIMD fashion. Each {\sc \textbf{Tulip}}-PE consists of a small network of \textit{binary neurons}, and a small amount of local memory per neuron. The unique aspect of the binary neuron is that it is implemented as a \textit{mixed-signal} circuit that natively performs the inner-product and thresholding operation of an artificial binary neuron. Moreover, the binary neuron, which is implemented as a single CMOS standard cell, is reconfigurable, and with a change in a single parameter, can implement all standard operations involved in a BNN. We present novel algorithms for mapping arbitrary nodes of a BNN onto the {\sc \textbf{Tulip}}-PEs. {\sc \textbf{Tulip}} was implemented as an ASIC in TSMC 40nm-LP technology.  To provide a fair comparison, a recently reported BNN that employs a conventional MAC-based arithmetic processor was also implemented in the same technology.  The results show that  {\sc \textbf{Tulip}} is consistently 3X more energy-efficient than the conventional design, without any penalty in performance, area, or accuracy. 
		\end{abstract}
		
		
		\section{Introduction}
		Convolutional (Deep) Neural Networks (CNNs or DNNs) have become a dominant algorithmic framework in machine learning due to their remarkable success in many diverse applications~\cite{deep-neural-networks-for-acoustic-modeling-in-speech-recognition,NIPS2012_4824,NIPS2015_5638,HUNT19921083,Liang_2015_CVPR}, etc., and even performing better than humans in some situations~\cite{He2015DelvingDI}.  DNNs are now being applied to domains that require compute-intensive operations performed on very large data sets, using models with millions of parameters~\cite{arxiv_VGG}. Consequently, extensive ongoing efforts are being made to improve their performance and energy efficiency.  
		
		Regardless of the hardware platform (CPU-GPU, FPGA, or ASIC) on which DNNs are deployed, the biggest challenge toward improving their performance and energy efficiency has been the on-chip storage requirement.  Cost and yield considerations limit the feasible on-chip storage to be one to two orders of magnitude smaller than what is required by many of the popular DNN models, forcing most of the parameters for even moderate size DNNs to be stored in off-chip DRAMs. This results in a large energy ($> 200X$) and delay ($> 10X$) penalties \cite{Nurvitadhi-FPT2016}.  This has accelerated efforts to drastically reduce the DRAM storage requirements and the associated access delays.  Some well-known methods include weight and synapse pruning, quantization (i.e. reducing bit widths of inputs and weight), weight sharing, Huffman coding, and approximate arithmetic, to name a few. 
		
		Quantization remains the most effective technique to reduce memory requirements and computation latency. An extreme form of quantization is to replace the weights and data by binary values, which results in drastic reductions in both storage requirements and computational latency. The resulting networks, known as binary neural networks (BNNs) \cite{Courbariaux_2016} have been shown to have nearly the same accuracy as DNNs on some small networks (MNIST, SVHN, and CIFAR10) \cite{Courbariaux_2016}, and similar accuracy to that of larger networks (AlexNet, GoogLeNet, ResNet) \cite{DBLP:journals/corr/RastegariORF16}.
		
		BNNs provide a good tradeoff between reduced energy consumption and improved performance against classification accuracy. As a result, they have generated sustained interest in the machine learning community, among researchers in VLSI architecture, circuits, and CAD, and leading  FPGA companies (i.e., Xilinx and Intel)~\cite{2017_Yaman_FINN_BNN,Anderson2017TheHG}. 
		
		A DNN is a directed acyclic graph (DAG), in which the nodes represent operations such as matrix-vector products, thresholding applied to inner-products, computation of the maximum of vectors, etc. In BNNs, such computations can be implemented almost entirely with binary operations.  This makes FPGAs a particularly practical choice for implementing BNNs.  Dedicated modules for each operation can be added to the design based on layer-specific requirements. These modules can be pipelined to maximize the throughput of the design. This approach amounts to mapping the nodes of the DAG, layer by layer, to corresponding modules on the FPGA.  Often, the entire BNN can be mapped onto the FPGA.  Examples of this design strategy, referred to as a \textit{dataflow} architecture, can be found in  \cite{2019_Yizing_NSTI,2018_Xiaoyu_XNOR-RRAM,2017_Yaman_FINN_BNN,2019_Tong_LP-BNN}.
		
		ASIC implementations of BNNs take a different approach.  In order to execute any BNN, their basic computational engine consists of a collection of processing elements (PEs), which are comprised of dedicated circuits to perform the operations specific to neural networks such as convolution, max-pooling, RELU, etc. Implementing a BNN on an ASIC next requires scheduling the execution of the nodes of the DAG on the PEs, while optimizing the intermediate storage and accesses to external memory.  This approach, referred to as a \textit{loopback} architecture, is the basis of many recent designs, for which examples may be found in \cite{2018_Andrawes_XNORBIN,2018_Renzo_YodaNN,2016_Hiroki_Mem_based_realization_BNN}.
		
		In this paper, we describe {\sc Tulip}, a new ASIC architecture to realize BNNs, designed with the aim of maximizing their energy efficiency. Although {\sc Tulip} falls under the category of a loopback architecture mentioned above, its processing element ({\sc Tulip}-PE) is radically different from the existing BNN accelerators, which leads to new algorithms to map BNNs onto {\sc Tulip}.   The key features of {\sc Tulip} and the contributions of this paper are summarized below.
		\begin{enumerate}
			\item {\sc Tulip} is a scalable SIMD machine, consisting of a collection of concurrently executing {\sc Tulip}-PEs.
			
			\item In addition to the design of {\sc Tulip}, this paper describes a new approach to map any BNN (any number of nodes and nodes with arbitrary fanin) onto {\sc Tulip}.
			
			\item The architecture of {\sc Tulip}-PE is radically different compared to the PEs in other BNN accelerators. It consists of a small, fully connected network of \textit{binary neurons} each with a small, fixed fanin.  A binary neuron is implemented as a \textit{mixed-signal} circuit that natively computes the inner-product and threshold operation of a neuron.  The mixed-signal binary neuron is implemented as a single \textit{standard cell}, that is just a little larger than a conventional flipflop. Moreover, the mixed-signal binary neuron is easily configured to perform \textit{all} the primitive operations required in a BNN. By suitably applying control inputs, a {\sc Tulip}-PE can be configured to perform all the operations required in a BNN, namely the accumulation of partial sums, comparison, max-pooling, and RELU operations.  Hence, exactly one such cell is needed to implement all necessary primitive functions in a BNN.
			
			\item Because the binary neurons within a {\sc Tulip}-PE have a fixed fanin, the function of a binary neuron with an arbitrarily large fanin has to be decomposed into a sequence of operations that have to scheduled on the {\sc Tulip}-PE.  A novel scheduling algorithm for this purpose is described.
			
			\item Due to the small area and delay of a single {\sc Tulip}-PE, several of these can be used within the same area that is occupied by a conventional MAC, and they can be operated in parallel. This, combined with the uniformity of the computation at the individual node and network levels, leads to significant improvement in energy efficiency, without sacrificing the area or performance.
		\end{enumerate}

		The paper is organized as follows: Section \ref{sec:background} describes a \textit{generic} architecture of a binary neuron, which is commonly referred to as a \textit{threshold gate}. Here, only the key characteristics of such an element are described and the details of the circuit design are omitted.  There are several recent publications describing the architecture of a threshold logic gate \cite{currentMode, Mozaffari_TNANO_2018, 2019_Wagle_ThresholdLogicInAFlash}, any one of which would be suitable for {\sc Tulip}. Section \ref{sec:decomposition} shows how the function of an arbitrarily large binary neuron can be efficiently decomposed into a computation tree consisting of smaller binary neurons that are mapped to the PEs of {\sc Tulip}. Section \ref{sec:architecture} describes how the novel PE is constructed and how it can be reconfigured to perform the various operations of the BNN. Section \ref{sec:results} compares the throughput, power, and area of {\sc Tulip} against the state of the art approaches. Finally, Section \ref{sec:conclusion} concludes this paper.
		
		\section{Background to Binary Neurons}
		\label{sec:background}
		
		A Boolean function $f(x_1, x_2, \cdots, x_n)$ is called a threshold function if there exist weights $w_i$ for $i = 1, 2, \cdots, n$ and a threshold $T$\footnote{\vspace{-8pt} W.L.O.G. the weights $w_i$ and threshold $T$ can be integers \cite{book:muroga}.} such that
		\begin{equation}
			\label{eq:thresholddef}
			f(x_1, x_2, \cdots x_n) = 1 ~\Leftrightarrow~ \sum_{i=1}^{n} w_i x_i \geq T,
		\end{equation}
		where $\sum$ denotes the arithmetic sum. Thus a threshold function can be represented as $(W, T) = [w_1, w_2, \cdots, w_n; T]$.  An example of a threshold function is $f (a,b,c,d) =      ab \lor ac \lor ad \lor bcd$, with  $[w_1, w_2, w_3, w_4; T] = [2, 1, 1, 1; 3]$.
		Threshold logic was first introduced by McCulloch and Pitts \cite{McCulloch_Pitts} in 1943 as a simple model of an artificial neuron. Since then, there has been an extensive body of work exploring the many theoretical and practical aspects of threshold logic \cite{book:muroga}. The recent resurgence of interest in neural networks has rekindled interest in threshold logic and its circuit realizations. A binary neuron is a threshold logic gate, and is therefore a circuit that realizes a threshold function.  Although there exist conventional static CMOS logic implementations of threshold functions, we do not use them in our work as they are inefficient in performance, power, and area. Instead, the binary neuron we consider in this paper is a mixed-signal implementation in which the defining inequality (Eq. \ref{eq:thresholddef}) is evaluated by directly comparing some electrical quantity such as charge, voltage or current \cite{Samuel_2010_ICM,currentMode}. Interest in binary neurons continues to grow with new architectures incorporating RRAMs, STT-MTJs, and flash transistors, demonstrating substantial improvements in performance, power, and area compared to their CMOS equivalents \cite{Mozaffari_TNANO_2018, 2019_Wagle_ThresholdLogicInAFlash} 
		
		\begin{figure}
			\centerline{\includegraphics[width=0.8\columnwidth]{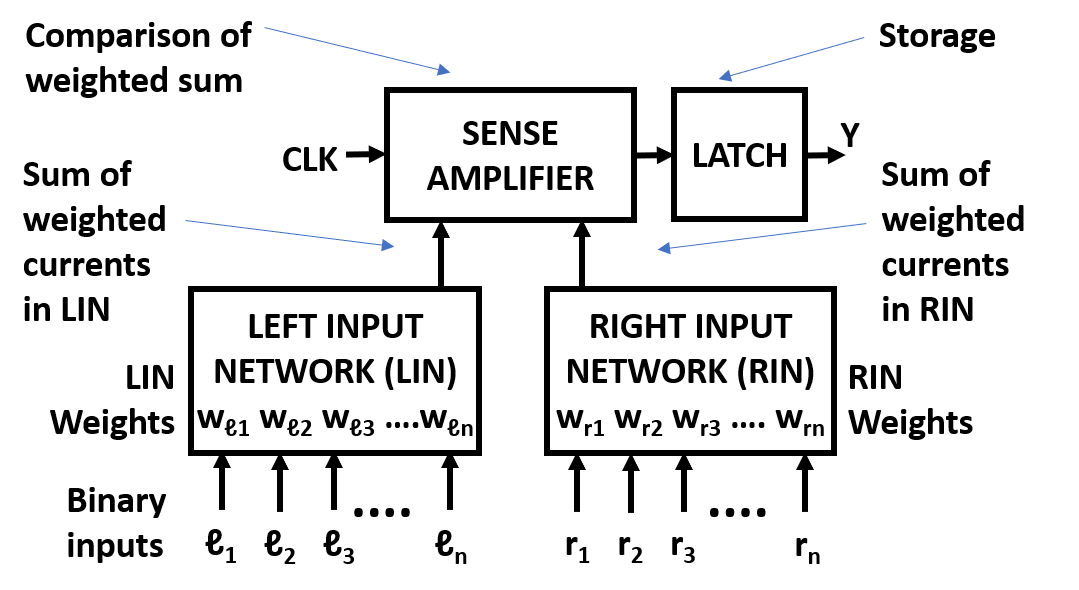}}
			\caption{\small Threshold Logic Gate (Neuron) Architecture }
			\label{fig:ftl_cell}
			\vspace{-15pt}
		\end{figure}
		
		Figure~\ref{fig:ftl_cell} shows an abstract block diagram of a circuit that serves as a template for nearly all the recent mixed-signal implementations of binary neurons \cite{Samuel_2010_ICM,2019_Wagle_ThresholdLogicInAFlash}.  It consists of four components: a left and right input network (LIN and RIN respectively), a sense amplifier, and a latch.  The key principle under which it operates is as follows.  The outputs of the sense amplifier are differential digital signals, with $(1,0)$ and $(0,1)$ setting and resetting the latch respectively.  The latch state remains unchanged when its inputs are $(0,0)$ or $(1,1)$. The weights $w_i$ that define the threshold function (Eq. \ref{eq:thresholddef}) are realized in ways that vary among different implementations \cite{Mozaffari_TNANO_2018, 2019_Wagle_ThresholdLogicInAFlash,currentMode}, but the common feature of all implementations is that they determine the charge, voltage or current of the LIN and RIN once the inputs are applied. That is, LIN and RIN are designed so that the charge, voltage, or current of the path that $x_i$ controls will be proportional to $w_i$.  The inputs $(x_1, x_2, \cdots, x_n)$ of a threshold function are mapped to the inputs of the LIN ($\ell_1, \ell_2, \cdots, \ell_n$) and RIN ($r_1, r_2, \cdots, r_n$) in \textit{such} a way that for every on-set (off-set) minterm, the charge, voltage or current of the LIN (RIN) reliably exceeds  that of the RIN (LIN) causing the sense amplifier to set (reset) the latch.  Ensuring that the inputs to the LIN and RIN are applied at a clock edge turns the circuit into a multi-input, edge-triggered flipflop, that computes the Boolean threshold function.

		\section{Binary Neural Network using Binary Neurons}
		\label{sec:decomposition}
		
		\begin{figure*}
			\small
			
			\centerline{\includegraphics[width=\linewidth]{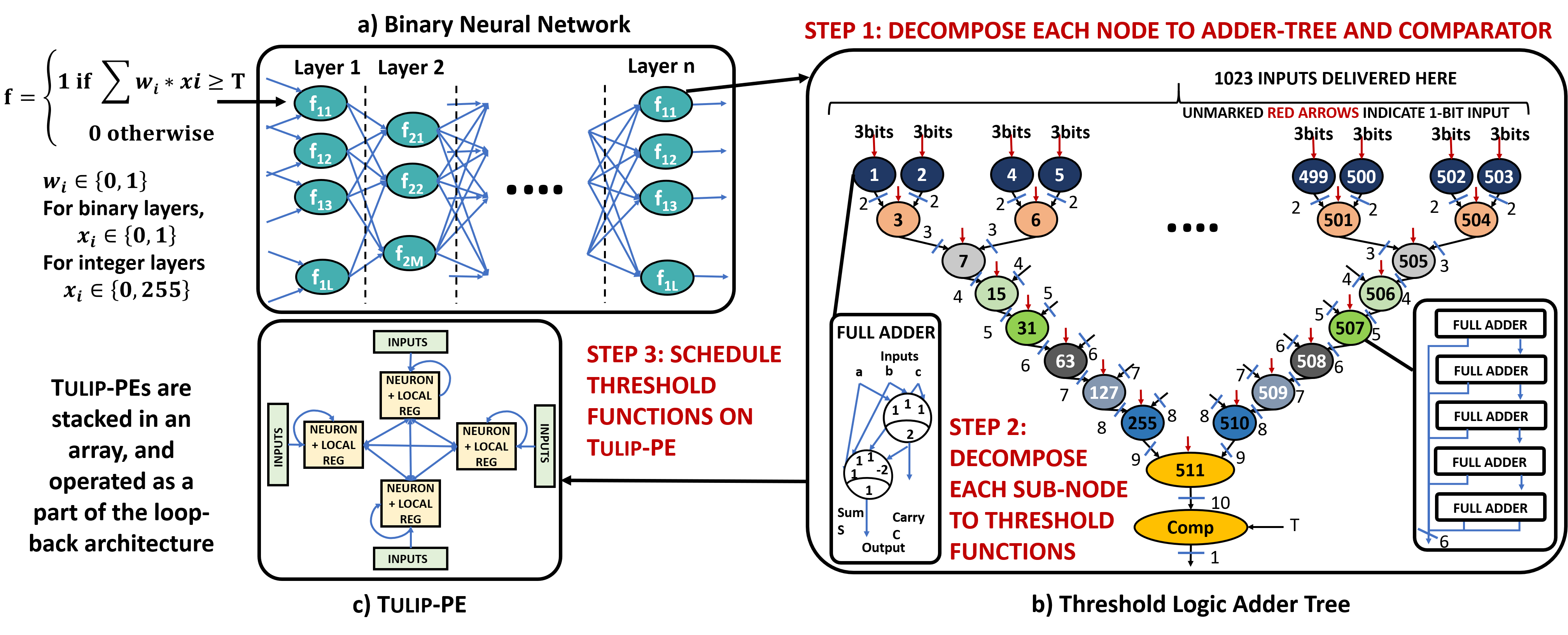}}
			\vspace{-5pt}
			\caption{\small {\sc Tulip} Flow: Each node of a BNN is decomposed into an adder-tree. Each sub-node of an adder-tree is decomposed into a network of two-level threshold functions. The decomposed network is scheduled using reverse post-order schedule (Indicated using node numbers; Unmarked red arrows indicate 1-bit input), on a {\sc Tulip}-PE built using a cluster of four hardware neurons. 
			}
			
			\label{fig:generic_bnn}
			\vspace{-16pt}
		\end{figure*} 
		
		A threshold function with a large number of inputs needs to be decomposed into a network (directed acyclic graph or DAG) of threshold functions with bounded fanin, each of which can be directly realized by a binary neuron.  Figure~\ref{fig:generic_bnn}(a) depicts such a network, in which each layer (level in the DAG) consists of a collection of threshold functions $f_{ij}$, where $i$ is the index of the layer and $j$ is the index of the function within any layer $i$.  The conventional approach taken by all recently reported BNN architectures is to \textit{accumulate} the partial sums (i.e. the LHS of inequality~\ref{eq:thresholddef}) using standard digital circuits using multiply and accumulate operations are performed. The final thresholding operation using a conventional binary comparator. This approach does not exploit the underlying special nature of the functions being computed, namely, the fact they are threshold functions. Another possible disadvantage of this approach is that it may use arithmetic operators of maximum width, regardless of how small the results of the partial sums are.\footnote{In general, adders of varying width may be utilized.}
		
		There are two basic approaches to decompose a given threshold function into a network of \textit{bounded fanin} threshold functions, several heuristic approaches \cite{Neutzling_ICCAD_2015,Kulkarni_TVLSI_2016} view the threshold function as any other logic function, and use existing logic synthesis tools to perform a technology-independent re-synthesis into a traditional logic network. This logic network is searched for subgraphs that are bounded-fanin threshold functions.  An exact and more elegant algorithm for this is presented in \cite{ANNAMPEDU201384}. It directly constructs a network of bounded-fanin threshold functions, in which each function performs thresholding on partial sums. Unfortunately, both these approaches result in extremely large networks.
		
		The architecture of {\sc Tulip} combines both the above described approaches in a novel way. Figure~\ref{fig:generic_bnn} depicts the design flow and the main components of {\sc Tulip}.  First, a BNN is expressed as a network of threshold functions $f_{ij}$ (Figure \ref{fig:generic_bnn}(a)).  Next, the LHS sum of each threshold function is decomposed into a tree of adders  (Figure \ref{fig:generic_bnn}(b)) of bounded size, and each such adder is realized by the repeated use of one configurable binary neuron (Figure \ref{fig:generic_bnn}(b), see insets).  This eliminates the waste incurred by conventional methods of accumulation that use operators of max-width. In Figure~\ref{fig:generic_bnn}(b) the labels inside the node show the order in which that node is executed on a {\sc Tulip}-PE for a threshold function with 1023 inputs. Note that although accumulation can be implemented by using conventional adders of varying sizes, the key difference with {\sc Tulip} is that \textit{all} the operations that arise in a BNN (addition, accumulation, comparison, and max-pooling) are implemented by the same, single configurable binary neuron in {\sc Tulip}.
		
		The main processing element in {\sc Tulip} ({\sc Tulip}-PE) consists of a complete network of 4 configurable binary neurons (as shown in Figure \ref{fig:generic_bnn}(c)).  The operations in the adder tree, as well as all the other operations in a BNN,  are scheduled to be executed on a {\sc Tulip}-PE so as to minimize the storage required for intermediate results. Each full-adder\footnote{This can be changed to implement a two-bit or three-bit carry-lookahead addition.  Doing so would simply require a binary neuron with a different set of weights, and could increase the throughput at the expense of a small increase in area and power. We plan to address this in future work.} is implemented as a cascade of two binary neurons (Figure ~\ref{fig:generic_bnn}(b), left inset). Larger width adders are implemented using a cascade of full adders (Figure ~\ref{fig:generic_bnn}(b) right inset).
		
		Finally, the top-level structure of {\sc Tulip} consists of a  number of PEs along with image and kernel buffers (Figure \ref{fig:tulip_top}). {\sc Tulip} is scalable, i.e., the throughout can simply be increased linearly by adding PEs and using larger image and kernel buffers, without changing the scheduling algorithm.

		\section{{\sc Tulip} Implementation}
		\label{sec:architecture}
		{\sc Tulip} involves the co-design and co-optimization of novel hardware and scheduler optimizations that together perform the operations of the BNN. In this section, the hardware architecture of the {\sc Tulip}-PE is discussed first. Then the scheduling algorithm needed to perform various operations such as addition, comparison, etc. is discussed. Finally, the top-level architecture is described, which uses an array of {\sc Tulip}-PEs to realize the entire BNN.
		
		\subsection{Hardware architecture of {\sc Tulip}-PE}
		\begin{figure}[h]
			\centerline{\includegraphics[width=0.95\columnwidth]{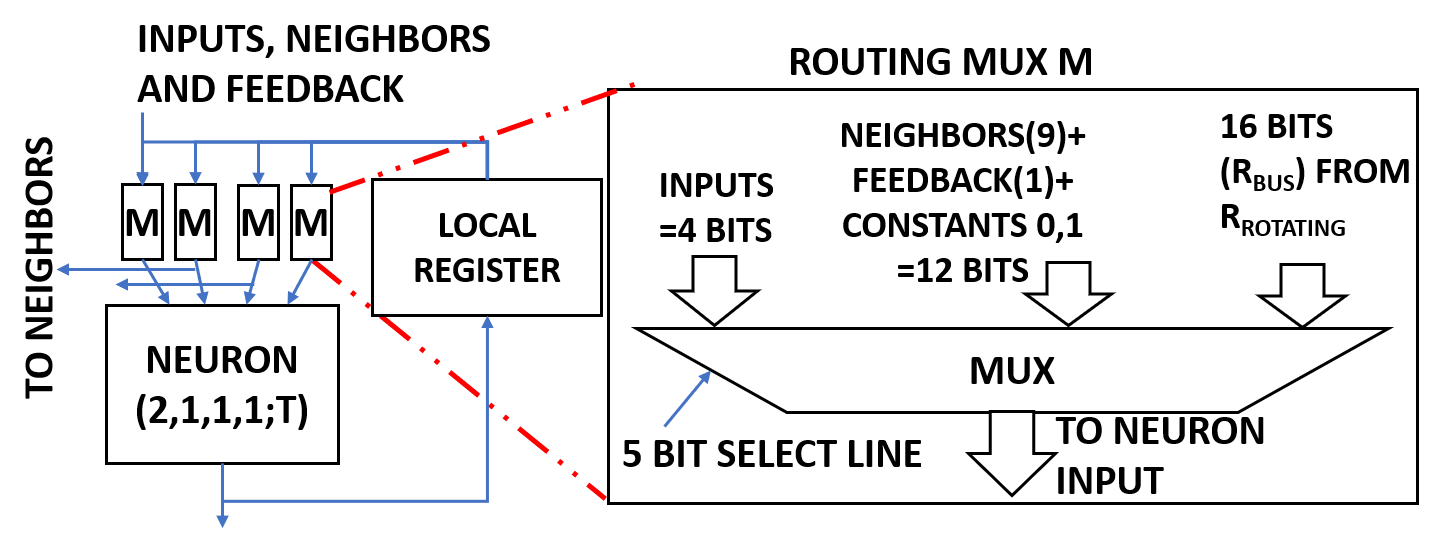}}
			\vspace{-5pt}
			\caption{\small The hardware neuron and its connections.}
			\label{fig:tulip_pe}
			\vspace{-18pt}
		\end{figure}
		
		A  {\sc Tulip}-PE (Figure \ref{fig:generic_bnn}(c)) has 4 fully connected neurons, referred to as $N1, \cdots, N4$,  each with 16-bit local registers. Each neuron of the {\sc Tulip}-PE is shown in Figure \ref{fig:tulip_pe}. Inter-neuron communication is implemented using multiplexers as shown in Figure \ref{fig:tulip_pe}. Each neuron has four inputs $a$, $b$, $c$, and $d$, with weights 2, 1, 1, and 1 respectively and a threshold T that is modified using digital control signals. The number of neurons in each {\sc Tulip}-PE is determined based on the computational requirements. The minimum number of neurons needed to perform addition, comparison, maxpooling, and RELU was found to be four, and was hence chosen for this paper. All 4 neurons of a {\sc Tulip}-PE share their inputs $b$ and $c$. This is done so that the neuron can fetch data from its local register, and broadcast it to all other neurons. The local registers are constructed using latches. As opposed to global registers, the local registers allow the neurons to access temporarily stored data faster, and also reduce the power consumption per read/write access.
		
		\begin{figure*}[ht!]{}
			\centering
			\includegraphics[width=\linewidth]{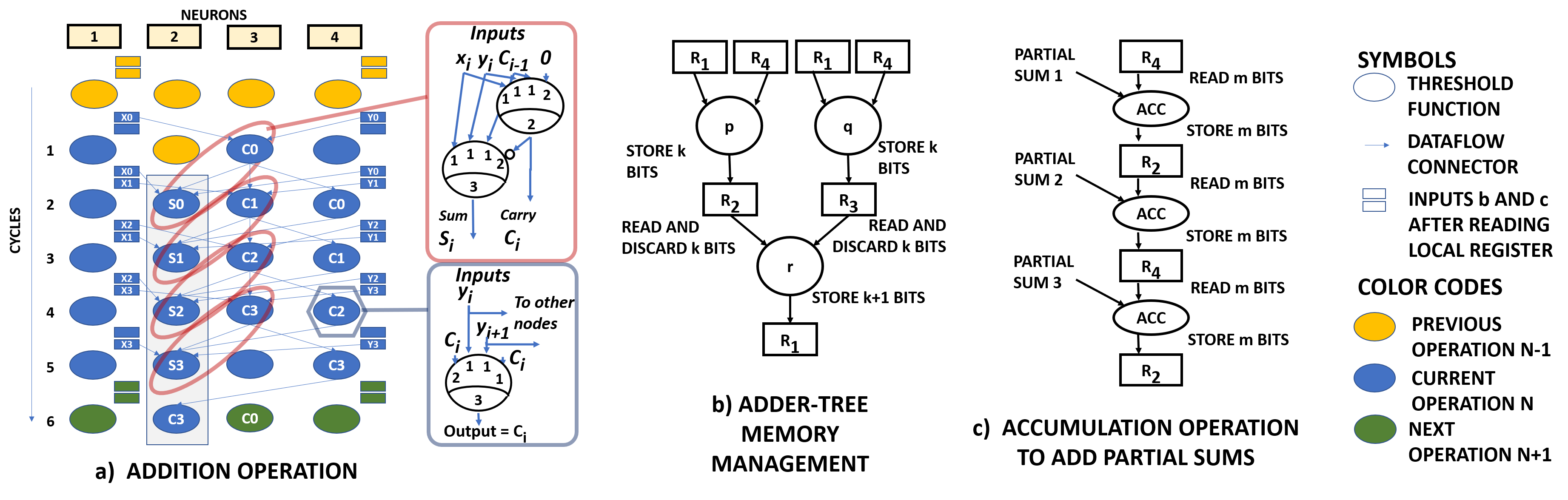}
			\vspace{-20pt}
			\caption{\small Adder, Adder-tree and Accumulator Schedule}
			\label{fig:all_sch_p1}
			\vspace{-15pt}
		\end{figure*}            
		
		\subsection{Decomposition and scheduling of an adder tree}
		In this section, we describe how a threshold function $f_{ij}$ in the BNN (Figure~\ref{fig:generic_bnn}(a)) is computed on a single {\sc Tulip}-PE (Figure~\ref{fig:generic_bnn} (b)). The node $f_{ij}$ computes the predicate $S \geq T$, where $S = \sum_{i} w_i x_i$.  The adder tree shown in Figure 2(b) is a binary decomposition of the $S$ into partial sums, with the leaf nodes (shown at the top) computing the sum of three inputs. The computation of partial sums uses a \textit{reverse post order} (RPO) scheme, which schedules the computation of a sum at a given node after both the sums associated with the left and right subtrees rooted at its left and right nodes have been computed. Therefore the number of bits required for the output of a node is one more than the number of bits of its inputs. In Figure~\ref{fig:generic_bnn}(b), the numeric label shown inside a node indicates its position in the RPO. The key property of the RPO is that it minimizes the maximum amount of storage required to store the intermediate results. 
		
		Consider the N-input adder tree shown in Figure \ref{fig:generic_bnn}b. The adder-tree has $\lfloor log_2(N)\rfloor$ levels, assuming that the leaf nodes are at level 0. Let $v$ be a node at level $i$ in the adder tree, and $v_l$ and $v_r$ be its left and right subtrees (both at level $i-1$). Let $m_i$ denote the maximum storage used for all computations up to and including a node at level $i$. Since the node at level $i$ corresponds to an $i+1$-input adder, the storage required for the output of a node at level $i$ is $i+2$. Since, the adder tree is balanced, without the loss of generality, we can assume $v_l$ is scheduled before $v_r$. To compute $v$, it is only required to store the output of $v_l$, which requires $i+1$ bits of storage. The maximum storage used to compute $v_l$ is $m_{i-1}$. Hence $m_i = i+1+m_{i-1}$, with $m_0=2$. Therefore, $m_i=(i^2+3i)/2+2$. As the highest level is $\lfloor log_2N\rfloor -1 $, the maximum required storage will be $(\lfloor log_2(N) \rfloor ^2 + \lfloor log_2(N)\rfloor)/2+1$. Therefore, an adder-tree has a storage requirement complexity of $O(log_2^2(N))$.

		\subsection{Addition and Accumulation Operation}
		For a node $p$ in the adder tree, assume neurons $N1$ and $N4$ broadcast two operands from $R_1$ and $R_4$, using the threshold function shown in Figure \ref{fig:all_sch_p1}(a) bottom-right inset. Then, $N2$ and $N3$ will be used to generate the sum and carry bits of $p$, over multiple cycles, using the threshold function shown in Figure \ref{fig:all_sch_p1}(a) top-right inset. Since the sum bits are computed on $N2$, the final result of $p$ will be stored in the local register of $N2$, i.e. $R_2$. Figure \ref{fig:all_sch_p1}(a) demonstrates the schedule for 4-bit addition (see node 15 of the adder-tree in Figure \ref{fig:generic_bnn}(b)) using two 4-bit operands $x$ and $y$, i.e. \{$x_3$,$x_2$,$x_1$,$x_0$\} and \{$y_3$,$y_2$,$y_1$,$y_0$\}. The final result of $x+y$ is stored in $R_2$.
		
		Now, consider nodes $p$, $q$, and $r$ in the adder-tree, as shown in Figure \ref{fig:all_sch_p1}(b). $r$ sums the results of  $p$ and $q$. Since the result of $p$ is stored in $R_2$, the result of $q$ is stored in $R_3$ to allow simultaneous reading of operands while computing $r$.  $r$ reads $R_2$ and $R_3$ to generate its sum bits on $N1$, and carry on $N4$. The memory used by the results of $p$ and $q$ can now be freed.
		Each addition operation stores its result to a specific memory location in the local registers so that the data in the memory is not prematurely overwritten during RPO scheduling.
		
		The adder-tree used in this paper handles up to 10-bit addition on a {\sc Tulip}-PE. However, this range can be further extended by configuring the {\sc Tulip}-PE for accumulation. Numbers can be added to an accumulated term stored in the local registers using a multi-cycle addition operation. Figure \ref{fig:all_sch_p1}(c) shows the addition of an input number $p$ with the accumulated term $q$. Since the same local register cannot provide the operands and store the results simultaneously, the storage of $q$ is alternated between the $R_2$ and $R_4$, for each new accumulation.

		\begin{figure*}[h]{}
			\centering
			\includegraphics[width=0.9\linewidth]{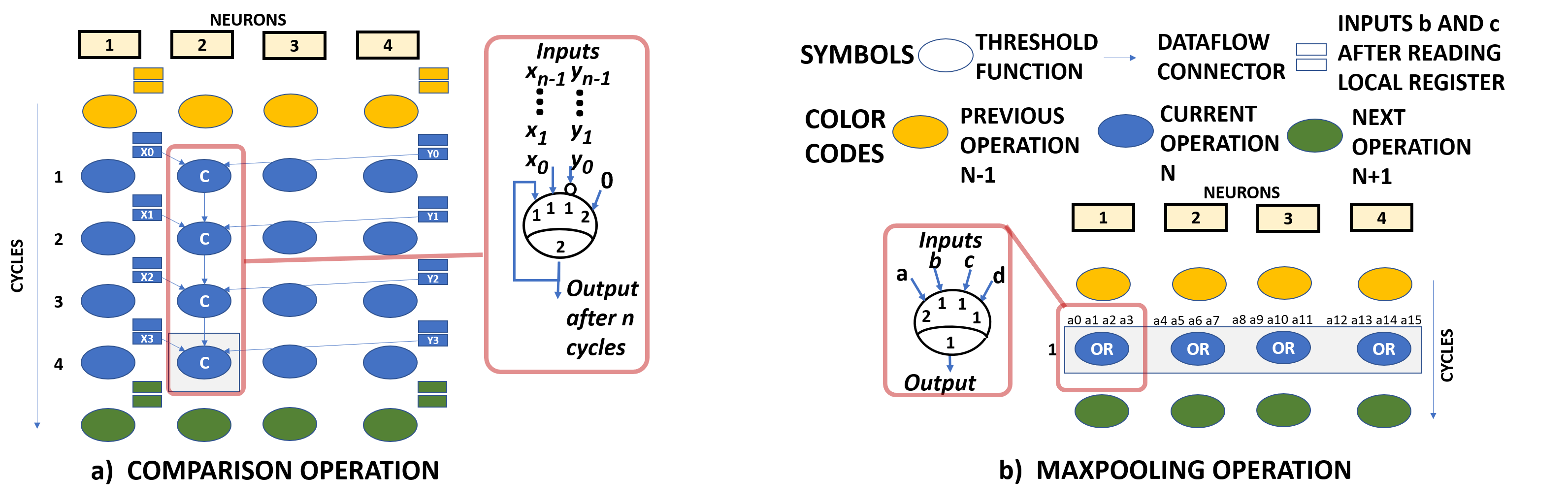}
			\vspace{-10pt}
			\caption{\small Comparator and Maxpooling Schedule}
			\label{fig:all_sch}
			\vspace{-15pt}
		\end{figure*}
		
		\subsection{Comparison, Batch Normalization, Maxpooling, RELU Operation}
		
		\textbf{Comparison:} A multi-cycle sequential comparator is implemented using 3-input threshold functions, as shown in Figure \ref{fig:all_sch}(a). To the best of our knowledge, this is the first implementation of a sequential comparator that uses 3-input neurons. Two $n$-bit numbers $x$ and $y$ that need to be compared are serially delivered from LSB to MSB to the comparator that returns the value of the predicate ($x>y$). In the first cycle, the LSBs of both numbers are compared. In the $i^{th}$ cycle of the comparison, if $x_i>y_i$, then the output is 1, and if $x_i<y_i$, then the output is 0. If $x_i=y_i$, then the result of the $(i-1)^{th}$ cycle is retained. The inset in Figure \ref{fig:all_sch}(a) shows the logic for bitwise comparison. At the end of $n$ cycles, the output is 1 if $x>y$, and 0 otherwise. The schedule of a 4-bit comparison is shown in Figure \ref{fig:all_sch}(a). The 4-bit inputs $x$ and $y$ are streamed to the comparator either through the local registers or through the input channels.
		
		\textbf{Batch Normalization:} This operation performs biasing of an input value in BNNs. For BNNs, it is realized by subtracting the value of bias from the threshold T of the binary neuron, as described in \cite{simons2019review}. Therefore, batch normalization in {\sc Tulip} is implemented using the comparison operation.
		
		\textbf{Maxpooling:} In a BNN, this operation is an OR operation on a pooling window of layer outputs. This can be implemented using the threshold gate shown in Figure \ref{fig:all_sch}(b). Each of the neurons implement one four-input OR function, without the need for local registers. The schedule for this operation requires a single cycle as shown in Figure \ref{fig:all_sch}(b).
		
		
		\textbf{RELU}: This implementation of RELU in {\sc Tulip} is also an extension of the comparator schedule shown above. In RELU, if the input value is greater than threshold T, then the output gets the value of the input, otherwise, it is 0. This is achieved by ANDing the result of the input value with the comparator's result, using a 2-input threshold function [1,1;2].

		\begin{figure}
			\centerline{\includegraphics[width=\columnwidth]{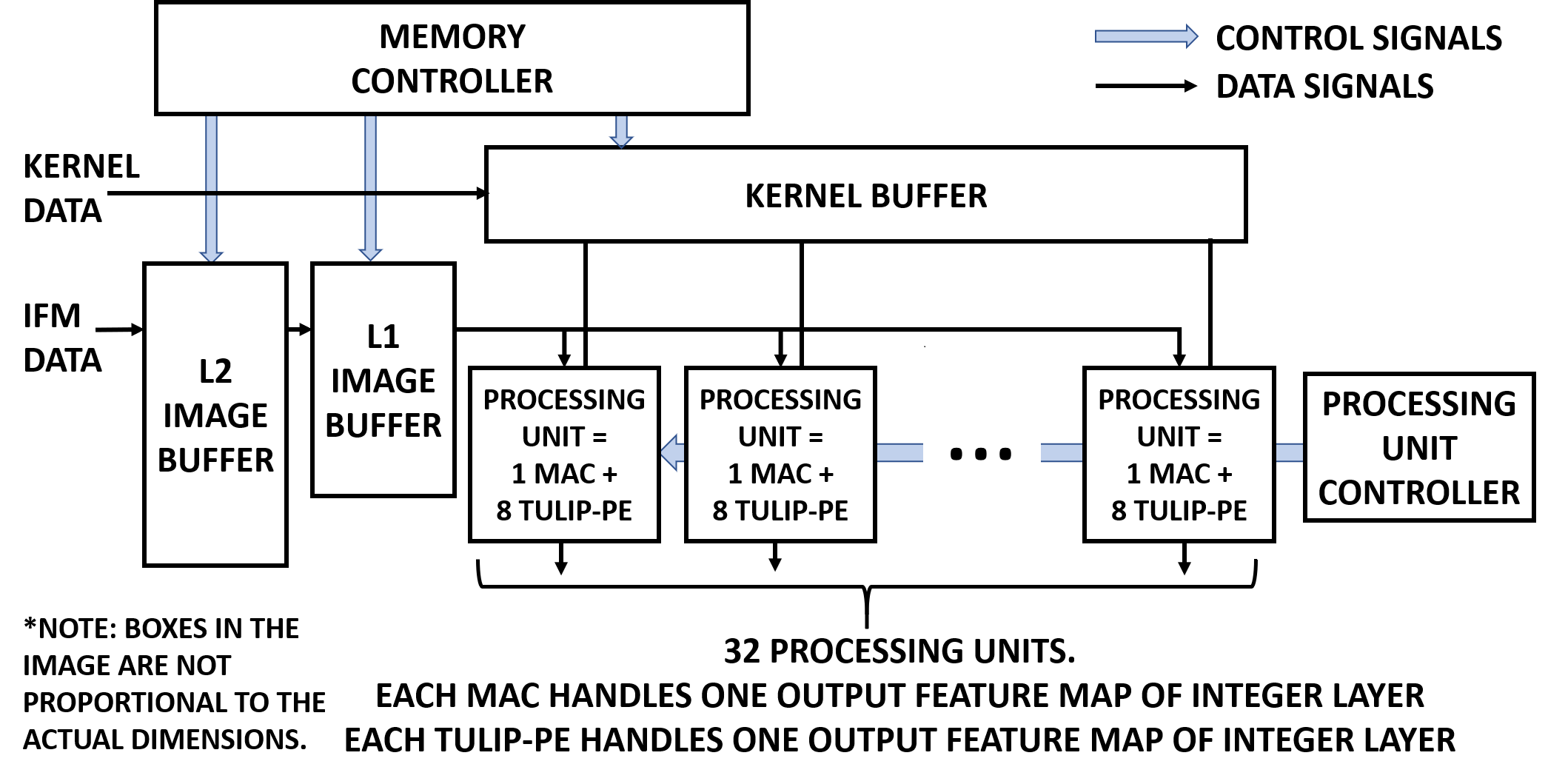}}
			
			\caption{\small {\sc Tulip} Top Level Architecture: Controller configures the processing units. Memory channels the input pixels and weights through image and kernel buffers. The output of the processing units is collected in the output buffers before sending back to the memory.}
			\label{fig:tulip_top}
			\vspace{-7pt}
		\end{figure}    
		\subsection{Top Level View of the Architecture}
		The top-level {\sc Tulip} architecture is shown in Figure \ref{fig:tulip_top}. It was designed to deliver high energy efficiency per operation while matching the throughput for the state-of-the-art implementations. This architecture consists of four major types of components: an image buffer, a kernel buffer, one or more processing units, and a controller. The kernel buffer is a shift-register which stores the weights of the BNN. Weights are populated on-chip before the inputs are loaded. The image buffer is a two-stage standard cell memory (SCM) named L2 and L1. Its use reduces off-chip communication (the SCM schedule is based on the technique presented in \cite{2018_Renzo_YodaNN}). In this architecture, 32 input feature maps (IFMs)\footnote{Memory can be scaled to store fewer or more IFMs} are loaded on-chip into L2 on a pixel-by-pixel basis. Once L2 is loaded with IFMs, L1 starts fetching the window of IFM pixels needed for the convolution operation, on a window-by-window basis.  This window of input pixels is broadcasted to all the processing units present in the design. The processing units are responsible for performing the convolution. These units also receive the appropriate weights from the kernel buffer. 
		
		A processing unit only triggers after necessary inputs and weights are received. The inputs and weights are multiplied using XNOR gates, to generate product terms. The processing unit has two components for accumulating the product terms: a MAC unit and eight {\sc Tulip}-PEs. A {\sc Tulip}-PE is used to handle an output feature map (OFM) of the binary layers. Although the {\sc Tulip}-PEs are capable of handling the integer layers as well, it would result in reduced throughput. This is because the {\sc Tulip}-PEs require several cycles for integer additions, which becomes progressively worse as the size of the operands increase. Hence, MACs are used for integer layers.
		
		The controllers used in the MAC units are simple counters. However, for the {\sc Tulip}-PEs, a reconfigurable sequence generator is used. This sequence generator follows the RPO schedule, and controls the local registers and the multiplexers of the {\sc Tulip}-PEs. The control signals are broadcast to all the processing units. The design of the controller is simple and has a negligible impact on the area and power of the overall {\sc Tulip} architecture. The {\sc Tulip} architecture also incorporates a clock gating strategy whenever a part of the design is not used. The necessary clock gating signals are also generated by the controller.
		
		Although the {\sc Tulip} architecture locks its configuration to a specific set of components for delivering weights and inputs, it can easily be tailored for a given application. For example, if a BNN does not have integer layers, then the MAC units can be removed, and the multi-bit input buffers can be trimmed to 1-bit input buffers. Various weight and input distribution techniques, such as the one presented in \cite{2016_Eyeriss} can also be used, by stacking the processing units in a 2-D arrangement instead of a 1-D configuration.

		\section{Experimental Results}
		\label{sec:results}
		\subsection{Experimental Setup}
		
		\begin{table}[]
			\small
			\centering
			\begin{tabular}{|c|c|c|c|}
				\hline
				& \begin{tabular}[c]{@{}c@{}}\textbf{Hardware}\\ \textbf{Neuron \cite{2019_Wagle_ThresholdLogicInAFlash}}\end{tabular} & \begin{tabular}[c]{@{}c@{}}\textbf{Logical}\\ \textbf{Equivalent}\end{tabular} & \begin{tabular}[c]{@{}c@{}}\textbf{X}\\ \textbf{Improve}\end{tabular} \\ \hline
				Area ($um^2$)     & 15.6 & 27     & 1.8X   \\ \hline
				Power($uW$)        & 4.46 & 6.72     & 1.5X  \\ \hline
				\begin{tabular}[c]{@{}c@{}}Worst Delay (ps)\end{tabular}   & 384 & 697       & 1.8X   \\ \hline
			\end{tabular}
			\caption{Hardware neuron versus standard cell neuron}
			\label{table:CMOS_vs_FTL}
			\vspace{-10pt}
		\end{table}

\begin{figure}[t]
			\centering
			\hspace{-120pt}
			\begin{subfigure}[t]{0.45\textwidth}
				\centerline{\includegraphics[width=0.4\columnwidth]{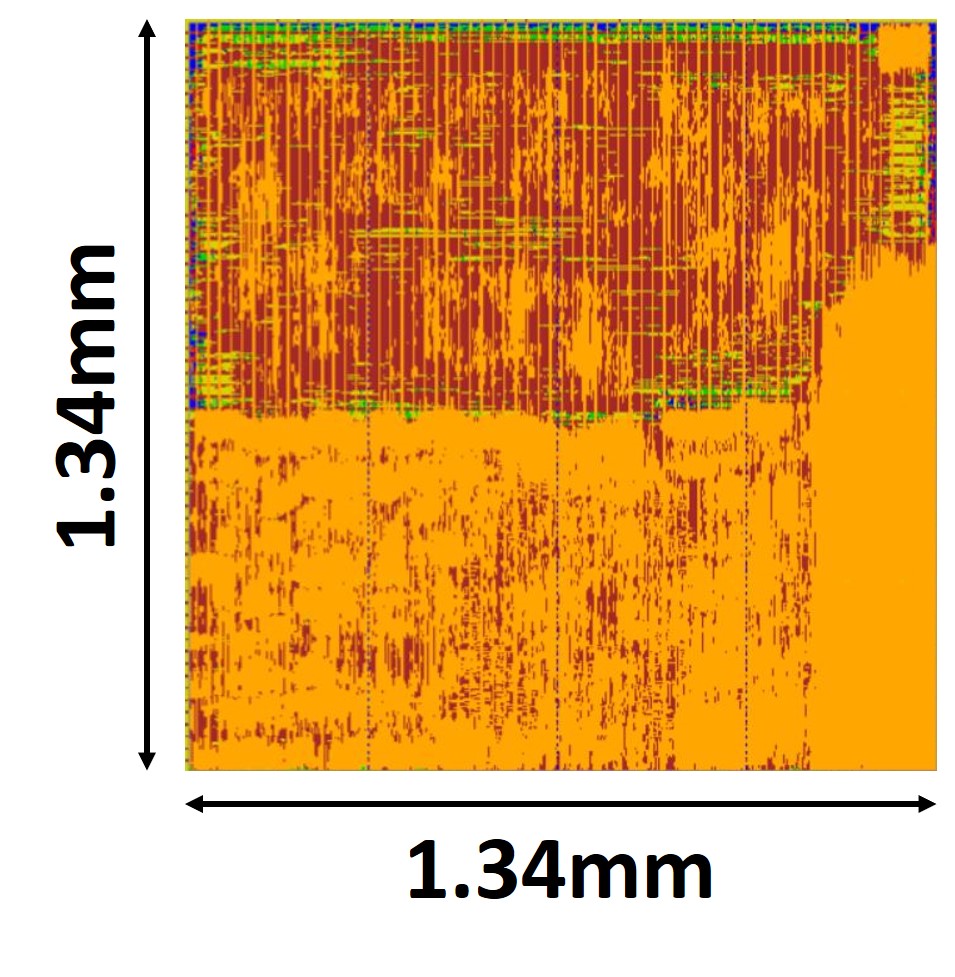}}
				\vspace{-15pt}
			\end{subfigure}%
			\hspace{-70pt}
			\begin{subfigure}{0.48\linewidth}
				\vspace{-84pt}
				\centering
				\small
				\scalebox{1}{\begin{tabular}{|c|c|}
						\hline
						Technology & TSMC 40LP   \\ \hline                    
						Area & 1.8 $mm^2$\\ \hline
						L2/L1/ & 680K/233K/  \\                     
						Kernel Area  &  468K $\mu m^2$  \\ \hline    
						Processing Unit. Area &  293K $\mu m^2$  \\ \hline
						Controller Area  &  4520 $\mu m^2$  \\ \hline                                        
						\# Std. Cells & 656K   \\ \hline
						\# Nets & 647K   \\ \hline
						Wirelength (m)           & 23.9 \\ \hline
						\# Metal Layers & 6 \\ \hline
					\end{tabular}
					
				}
			\end{subfigure}
			\caption{\small Layout of {\sc Tulip} Architecture in TSMC 40nm-LP}
			\label{fig:tulip_pnr}
			\vspace{-10pt}
		\end{figure}

		The {\sc Tulip} architecture was built based on the hardware neuron described in \cite{2019_Wagle_ThresholdLogicInAFlash}. The neuron was re-implemented in a 40nm technology, programmed to [2,1,1,1;T], and characterized across corners ($SS~0.81V~125^\circ C$, $TT~0.9V~25^\circ C$ and $FF~0.99V~0^\circ C$). The value of T is switched during run-time by changing the appropriate control signals of the neuron. Table \ref{table:CMOS_vs_FTL} demonstrates that this hardware neuron is substantially better than its conventional CMOS standard cell equivalent in terms of area, power, and delay. This is significant since {\sc Tulip} uses this neuron for all operations (computation of partial sums, comparison, RELU, and maxpool). {\sc Tulip} was synthesized and placed using TSMC 40nm-LP standard cells with Cadence Genus\textsuperscript{\textcopyright} and Innovus\textsuperscript{\textcopyright} (Figure \ref{fig:tulip_pnr}). The VCD file generated using real BNN workloads was used for power analysis, to model switching activity accurately.
		
		We compare {\sc Tulip} with a recent BNN design named YodaNN \cite{2018_Renzo_YodaNN} which was designed in 65nm UMC technology. To make a fair comparison, we implemented the entire YodaNN design in the same technology as {\sc Tulip}(40nm-LP from TSMC), and synthesized, placed and routed both the designs. Both the {\sc Tulip} and YodaNN were designed for up to 12-bit inputs, with binary weights. Therefore, for YodaNN, we added clock gating for 11/12 input bits when binary layers are evaluated. There are other ASIC architectures available, such as XNORBIN \cite{2018_Andrawes_XNORBIN}, which use more advanced memory techniques to improve energy efficiency. However, these architectures do not support integer layers and are therefore not suitable for comparison. Although \cite{2018_Renzo_YodaNN} does not report the throughput and energy efficiency for fully connected layers, we estimate the throughput and power by performing an element-wise matrix multiplication using the MAC units present in their architecture.

		\begin{table}[]
			\small
			\centering
			\begin{tabular}{|c|c|c|c|}
				\hline
				\textbf{\begin{tabular}[c]{@{}c@{}}Single\\   PE Metrics\end{tabular}} & \textbf{\begin{tabular}[c]{@{}c@{}}YodaNN MAC\\   (B)\end{tabular}} & \textbf{\begin{tabular}[c]{@{}c@{}}{\sc \textbf{Tulip}}-PE\\  (T)\end{tabular}} & \textbf{\begin{tabular}[c]{@{}c@{}}Ratio(X)\\(B/T)\end{tabular}} \\ \hline
				Area(um\textasciicircum{}2)  & 3.54E+04 & 1.53E+03       & 23.18  \\ \hline
				Power(mW)      & 7.17 & 0.12       & 59.75   \\ \hline
				Cycles  & 17 & 441       & 0.038     \\ \hline
				\begin{tabular}[c]{@{}c@{}}Time period(ns)\end{tabular}     & 2300 & 2300       & 1    \\ \hline
				Time(ns)       & 39 & 1014       & 0.038   \\ \hline
			\end{tabular}
			\caption{Comparison of fully reconfigurable MAC unit \cite{2018_Renzo_YodaNN} with a {\sc Tulip}-PE, for a 288 input neuron (Kernel =3x3)}
			\label{table:mac_vs_tulip_module}
			\vspace{-10pt}
		\end{table}

		\subsection{Evaluation of {\sc Tulip}-PE against MAC}        
		In Table \ref{table:mac_vs_tulip_module}, the 15-bit  reconfigurable MAC unit based on the design present in YodaNN\cite{2018_Renzo_YodaNN} is compared against the {\sc Tulip}-PE module. The MAC unit used in YodaNN is capable of handling 3x3, 5x5 and 7x7 kernel sizes. Note that both the MAC unit and {\sc Tulip}-PE are capable of handling integer inputs and binary weights. In large BNN architectures such as Alexnet\cite{xnornet}, the initial layers are integer layers, while the rest of the layers are binary. YodaNN uses MAC units for all layers while {\sc Tulip} uses {\sc Tulip}-PEs for binary layers and simplified MACs (which support only 5x5 and 7x7 kernel windows) for integer layers. Since the computation technique between YodaNN and {\sc Tulip} differs only for binary layers, the comparison of the MAC and the {\sc Tulip} is done for the binary layers. That is, both modules perform the weighted sum for binary activations and binary weights of 288 inputs, i.e. 3x3 kernel for 32 IFMs. Based on the Table \ref{table:mac_vs_tulip_module}, we note that the {\sc Tulip}-PE is 23.18X smaller than the MAC unit and consumes 60X less power. However, it consumes 27X more time as compared to the MAC unit, since it performs bit-level addition. The power delay product of a {\sc Tulip}-PE is 2.27X lower than the MAC unit, while at the same time being 23X smaller than the MAC.
		
		The use of an adder-tree based schedule helps the {\sc Tulip}-PEs deliver a better power-delay product than a conventional MAC unit. Furthermore, since a MAC unit is not capable of operations such as comparison, maxpooling, etc., the data is sent to other parts of the chip for these operations in \cite{2018_Renzo_YodaNN}. However, the {\sc Tulip}-PE, is capable of preserving the data locality and can perform the comparison and max-pooling operations internally, without the need to move the data to other modules, which saves additional energy.

		\subsection{Evaluation of the {\sc Tulip} Architecture}
		The following notation is used for evaluating the {\sc Tulip} architecture. For 2-D convolution, let ($x_{1},y_{1},z_{1}$) and ($x_{2},y_{2},z_{2}$) denote the dimensions of the IFMs and OFMs respectively. Let the kernel window size be $(k\times k)$.
		
		The number of processing units in {\sc Tulip} can be scaled to suit the application. However, for the sake of evaluation, {\sc Tulip} was designed with 32 simplified MAC units and 256 {\sc Tulip}-PEs, to ensure that the chip area of Tulip matches that of YodaNN. Note that the simplified MAC unit is not reconfigurable, and hence consumes significantly lower area and power than the MAC presented in YodaNN. Therefore, for {\sc Tulip}, convolution in done in batches of 32 OFMs at a time for integer layers, and 256 OFMs at a time for binary layers. Since the IFMs are re-fetched for each batch of OFMs, they are fetched Z = $z_{2}$/32 times for integer layers and $z_{2}$/256 times for binary layers. The YodaNN architecture uses 32 fully reconfigurable MAC units, and occupies the same area as {\sc Tulip}. Therefore, the number of times YodaNN fetches IFMs (Z) = $z_{2}$/32. Additionally, when the kernel size is small ($k \le 5$), the MAC units in both the designs can fetch twice the number of IFMs. Since the {\sc Tulip} can initiate more OFMs for binary layers, it significantly reduces the number of times an input needs to be fetched. For this paper, both the YodaNN and {\sc Tulip} architecture load 32 IFMs at a time on-chip. This specification can however be changed to meet the application requirements. If the total IFMs cannot fit on-chip, the OFMs are generated in pieces of P partial results. These partial results are later accumulated on-chip to generate the final OFM. For both the architectures in this paper, P = $z_{1}$/32. The total number of operations is counted by considering addition and multiplication separately. For a 2-D convolution layer, the total multiply and accumulate operations in {\sc Tulip} are $2z_{1}k^2x_{2}y_{2}z_{2}$, and for comparison of each accumulated sum with T, it is $x_{2}y_{2}z_{2}$.    For Alexnet, Table \ref{table:fetch_rqmt} compares the number of times the inputs need to be refetched (Z), and the number of times the P partial products need to be computed for both YodaNN and {\sc Tulip}. Since both the designs use MAC units for integer layers, there is no difference in both P and Z. However, for binary layers, {\sc Tulip} demonstrates 3X to 4X improvement in overall input-refetch (indicated by P$\times$Z) as compared to the YodaNN architecture.   
		
		
		
		Table \ref{table:conv_compare} and Table \ref{table:label_compare} compare the characteristics of YodaNN with {\sc Tulip}. Table  \ref{table:conv_compare} presents the results for the convolution layers and Table \ref{table:label_compare} presents the results for the entire BNN. The TULIP architecture outperforms YodaNN in energy efficiency by about 3X for the convolution layers. This is due to the combined use of adder-tree based schedule, coupled with clock gating. The energy efficiency also increases due to better input re-use, which allows the throughput to improve slightly. Considering all layers, {\sc Tulip}'s energy efficiency is 2.4X better than YodaNN. This is because memory consumes significantly more energy than the processing units when executing fully connected layers, which slightly diminishes the energy efficiency achieved in the convolution layers. The results also show that the gains are consistent across different neural networks.
		
		\begin{table}[]
			\small
			\centering
			\begin{tabular}{|c|c||c|c|c||c|c|c|}
				\hline
				\multirow{2}{*}{\textbf{\begin{tabular}[c]{@{}c@{}}Convolution\\ Layers\end{tabular}}} & \multirow{2}{*}{\textbf{Parts}} & \multicolumn{3}{c||}{\textbf{YodaNN}}   & \multicolumn{3}{c|}{\textbf{{\sc \textbf{Tulip}}}}    \\ \cline{3-8} 
				&     & \textbf{P} & \textbf{Z} & \textbf{P*Z} & \textbf{P} & \textbf{Z} & \textbf{P*Z} \\ \hline
				1 (Integer)      & 4   & 1   & 3   & 3     & 1   & 3   & 3     \\ \hline
				2 (Integer)      & 1   & 2   & 8   & 16    & 2   & 8   & 16    \\ \hline
				3 (Binary)       & 1   & 4   & 12  & 48    & 8   & 2   & 16    \\ \hline
				4 (Binary)       & 1   & 6   & 12  & 72    & 12  & 2   & 24    \\ \hline
				5 (Binary)       & 1   & 6   & 8   & 48    & 12  & 1   & 12    \\ \hline
			\end{tabular}
			\caption{Effect of input fetch requirements based on Alexnet layers for YodaNN and {\sc Tulip}. P: Number of times partial products are computed. Z: Number of times inputs are fetched into L2 and L1 buffers for OFM calculation.}
			\label{table:fetch_rqmt}
			\vspace{-10pt}
		\end{table}

		\begin{table}[]
			\small
			\centering
			\begin{tabular}{|c||c|c||c|c|}
				\hline
				\textbf{Conv only}  & \multicolumn{2}{c||}{\textbf{BinaryNet}}  &     \multicolumn{2}{c|}{\textbf{\textbf{AlexNet}}}  \\ \hline
				\textbf{Dataset}    & \multicolumn{2}{c||}{\textbf{\textbf{CIFAR10}}}     & \multicolumn{2}{c|}{\textbf{\textbf{Imagenet}}}      \\ \hline
				& YodaNN    & {\sc Tulip} (X) & YodaNN   & {\sc Tulip}  (X) \\ \hline
				Op.(MOp)   & 1017    & 1017 (1.0) & 2050   & 2050 (1.0) \\ \hline
				\begin{tabular}[c]{@{}c@{}}Perf.(GOp/s)\end{tabular}     & 47.6      & 49.5  (1.0) & 72.9     & 79.1   ( 1.1) \\ \hline
				\begin{tabular}[c]{@{}c@{}}Energy(uJ)\end{tabular}       & 472.6     & 159.1   (3.0) & 678.8    & 224.5   (3.0) \\ \hline
				\begin{tabular}[c]{@{}c@{}}Time(ms)\end{tabular} & 21.4      & 20.6    (1.0) & 28.1     & 25.9    (1.1) \\ \hline
				\begin{tabular}[c]{@{}c@{}}En.Eff.\\(TOp/s/W)\end{tabular} & 2.2       & 6.4     (3.0) & 3.0      & 9.1     (3.0) \\ \hline
			\end{tabular}
			\caption{Comparison of YodaNN with {\sc Tulip} architecture for accelerating convolution layers of standard datasets.}
			\label{table:conv_compare}
			\vspace{-7pt}
		\end{table}

		\begin{table}[]
			\small
			\begin{tabular}{|c||c|c||c|c|}
				\hline
				\textbf{All Layers}  & \multicolumn{2}{c||}{\textbf{BinaryNet}}  &     \multicolumn{2}{c|}{\textbf{\textbf{AlexNet}}}  \\ \hline
				\textbf{Dataset}    & \multicolumn{2}{c||}{\textbf{\textbf{CIFAR10}}}     & \multicolumn{2}{c|}{\textbf{\textbf{Imagenet}}}      \\ \hline
				& YodaNN    & {\sc Tulip}  (X) & YodaNN   & {\sc Tulip}  (X) \\ \hline
				Op.(MOp) & 1036    & 1036  (1.0) & 2168   & 2168  (1.0) \\ \hline
				\begin{tabular}[c]{@{}c@{}}Perf.(GOp/s)\end{tabular}     & 37.7      & 35.8    (0.9) & 12.3     & 13.1  ( 1.1) \\ \hline
				\begin{tabular}[c]{@{}c@{}}Energy(uJ)\end{tabular}       & 495.2     & 183.9   (2.7) & 1013.3   & 427.5  (2.4) \\ \hline
				\begin{tabular}[c]{@{}c@{}}Time(ms)\end{tabular} & 27.5      & 28.9   (0.9) & 176.8    & 165.0  (1.1) \\ \hline
				\begin{tabular}[c]{@{}c@{}}En.Eff.\\(TOp/s/W)\end{tabular} & 2.1       & 5.6    (2.7) & 2.1      & 5.1    (2.4) \\ \hline
			\end{tabular}
			\caption{Comparison of YodaNN with {\sc Tulip} for accelerating entire BNNs of standard datasets }
			\label{table:label_compare}
			\vspace{-10pt}
		\end{table}

		\section{Conclusion}
		\label{sec:conclusion}
		This paper is the first implementation of {\sc Tulip}, a BNN accelerator that uses current-mode binary neurons, and demonstrates up to 3X improvement in energy efficiency against a state of the art BNN hardware accelerator \cite{2018_Renzo_YodaNN}, without using the standard low power techniques such as voltage scaling and approximate computing. The {\sc Tulip} design uses the same area as \cite{2018_Renzo_YodaNN}, and slightly improves the throughput. The gains are achieved because {\sc Tulip} uses an adder-tree based schedule, instead of an accumulator. The gains are further boosted through the use of processing elements ({\sc Tulip}-PEs) built using a special arrangement of hardware neurons.  These {\sc Tulip}-PEs have extremely low area and power footprint compared to the existing realizations of the same function. As a result, {\sc Tulip} can deploy an order of magnitude more PEs as compared to a MAC-based architecture for the same chip area.
		\bibliography{TULIP_arxiv}

\begin{thebibliography}{10}

\bibitem{deep-neural-networks-for-acoustic-modeling-in-speech-recognition}
G.~Hinton, L.~Deng, D.~Yu, G.~Dahl, A.R. Mohamed, N.~Jaitly, A.~Senior,
  V.~Vanhoucke, P.~Nguyen, B.~Kingsbury, and T.~Sainath.
\newblock Deep neural networks for acoustic modeling in speech recognition.
\newblock {\em IEEE Signal Processing Magazine}, 29:82--97, November 2012.

\bibitem{NIPS2012_4824}
A.~Krizhevsky, I.~Sutskever, and G.~Hinton.
\newblock Imagenet classification with deep convolutional neural networks.
\newblock In F.~Pereira, C.~J.~C. Burges, L.~Bottou, and K.~Q. Weinberger,
  editors, {\em Advances in Neural Information Processing Systems 25}, pages
  1097--1105. Curran Associates, Inc., 2012.

\bibitem{NIPS2015_5638}
S.~Ren, K.~He, R.~Girshick, and J.~Sun.
\newblock Faster {R-CNN}: Towards real-time object detection with region
  proposal networks.
\newblock In C.~Cortes, N.~D. Lawrence, D.~D. Lee, M.~Sugiyama, and R.~Garnett,
  editors, {\em Advances in Neural Information Processing Systems 28}, pages
  91--99. Curran Associates, Inc., 2015.

\bibitem{HUNT19921083}
K.J. Hunt, D.~Sbarbaro, R.~Zbikowski, and P.~Gawthrop.
\newblock Neural networks for control systems survey.
\newblock {\em Automatica}, 28(6):1083 -- 1112, 1992.

\bibitem{Liang_2015_CVPR}
Ming Liang and Xiaolin Hu.
\newblock Recurrent convolutional neural network for object recognition.
\newblock In {\em The IEEE Conference on Computer Vision and Pattern
  Recognition (CVPR)}, June 2015.

\bibitem{He2015DelvingDI}
K.~He, X.~Zhang, S.~Ren, and J.~Sun.
\newblock Delving deep into rectifiers: Surpassing human-level performance on
  imagenet classification.
\newblock {\em 2015 IEEE International Conference on Computer Vision (ICCV)},
  pages 1026--1034, 2015.

\bibitem{arxiv_VGG}
K.~Simonyan and A.~Zisserman.
\newblock Very deep convolutional networks for large-scale image recognition.
\newblock 2014.

\bibitem{Nurvitadhi-FPT2016}
E.~Nurvitadhi, G.~Venkatesh, J.~Sim, D.~Marr, R.~Huang, J.~Ong Gee~Hock, Y.~T.
  Liew, K.~Srivatsan, D.~Moss, S.~Subhaschandra, and G.~Boudoukh.
\newblock {Can FPGAs Beat GPUs in Accelerating Next-Generation Deep Neural
  Networks?}
\newblock In {\em Proceedings of the 2017 ACM/SIGDA}, FPGA '17, pages 5--14,
  New York, NY, USA, 2017. ACM.

\bibitem{Courbariaux_2016}
M.~Courbariaux, I.~Hubara, D.~Soudry, R.~El-Yaniv, and .~Bengio.
\newblock Binarized neural networks: Training deep neural networks with weights
  and activations constrained to +1 or -1.
\newblock 2016.

\bibitem{DBLP:journals/corr/RastegariORF16}
M.~Rastegari, V.~Ordonez, J.~Redmon, and A.~Farhadi.
\newblock Xnor-net: Imagenet classification using binary convolutional neural
  networks.
\newblock {\em CoRR}, abs/1603.05279, 2016.

\bibitem{2017_Yaman_FINN_BNN}
Yaman Umuroglu, Nicholas~J. Fraser, Giulio Gambardella, Michaela Blott, Philip
  Leong, Magnus Jahre, and Kees Vissers.
\newblock {FINN: A Framework for Fast, Scalable Binarized Neural Network
  Inference}.
\newblock In {\em Proceedings of the 2017 ACM/SIGDA International Symposium on
  Field-Programmable Gate Arrays}, FPGA ’17, pages 65--74, New York, NY, USA,
  2017. Association for Computing Machinery.

\bibitem{Anderson2017TheHG}
A.~G. Anderson and C.~P. Berg.
\newblock The high-dimensional geometry of binary neural networks.
\newblock {\em CoRR}, abs/1705.07199, 2017.

\bibitem{2019_Yizing_NSTI}
Y.~{Li}, Z.~{Liu}, W.~{Liu}, Y.~{Jiang}, Y.~{Wang}, W.~L. {Goh}, H.~{Yu}, and
  F.~{Ren}.
\newblock {A 34-FPS 698-GOP/s/W Binarized Deep Neural Network-Based Natural
  Scene Text Interpretation Accelerator for Mobile Edge Computing}.
\newblock {\em IEEE Transactions on Industrial Electronics}, 66(9):7407--7416,
  2019.

\bibitem{2018_Xiaoyu_XNOR-RRAM}
X.~{Sun}, S.~{Yin}, X.~{Peng}, R.~{Liu}, J.~{Seo}, and S.~{Yu}.
\newblock {XNOR-RRAM: A scalable and parallel resistive synaptic architecture
  for binary neural networks}.
\newblock In {\em 2018 Design, Automation Test in Europe Conference Exhibition
  (DATE)}, pages 1423--1428, 2018.

\bibitem{2019_Tong_LP-BNN}
T.~{Geng}, T.~{Wang}, C.~{Wu}, C.~{Yang}, S.~L. {Song}, A.~{Li}, and
  M.~{Herbordt}.
\newblock {LP-BNN: Ultra-low-Latency BNN Inference with Layer Parallelism}.
\newblock In {\em 2019 IEEE ASAP}, volume 2160-052X, pages 9--16, 2019.

\bibitem{2018_Andrawes_XNORBIN}
A.~{Al Bahou}, G.~{Karunaratne}, R.~{Andri}, L.~{Cavigelli}, and L.~{Benini}.
\newblock {XNORBIN: A 95 TOp/s/W hardware accelerator for binary convolutional
  neural networks}.
\newblock In {\em 2018 IEEE Symposium in Low-Power and High-Speed Chips (COOL
  CHIPS)}, pages 1--3, 2018.

\bibitem{2018_Renzo_YodaNN}
R.~Andri, L.~Cavigelli, D.~Rossi, and L.~Benini.
\newblock {YodaNN: An Architecture for Ultralow Power Binary-Weight CNN
  Acceleration}.
\newblock {\em IEEE Transactions on Computer-Aided Design of Integrated
  Circuits and Systems}, PP:1--1, March 2017.

\bibitem{2016_Hiroki_Mem_based_realization_BNN}
H.~{Nakahara}, H.~{Yonekawa}, T.~{Sasao}, H.~{Iwamoto}, and M.~{Motomura}.
\newblock {A memory-based realization of a binarized deep convolutional neural
  network}.
\newblock In {\em 2016 International Conference on Field-Programmable
  Technology (FPT)}, pages 277--280, 2016.

\bibitem{currentMode}
S.~Bobba and I.~Hajj.
\newblock {Current-Mode Threshold Logic Gates}.
\newblock In {\em Proc. of ICCD}, pages 235--240, 2000.

\bibitem{Mozaffari_TNANO_2018}
S.~N. {Mozaffari} and S.~{Tragoudas}.
\newblock Maximizing the number of threshold logic functions using resistive
  memory.
\newblock {\em IEEE TNANO}, 17(5), Sep. 2018.

\bibitem{2019_Wagle_ThresholdLogicInAFlash}
A.~{Wagle}, G.~{Singh}, J.~{Yang}, S.~{Khatri}, and S.~{Vrudhula}.
\newblock Threshold logic in a flash.
\newblock In {\em 2019 IEEE 37th International Conference on Computer Design
  (ICCD)}, pages 550--558, 2019.

\bibitem{book:muroga}
S.~Muroga.
\newblock {\em { Threshold Logic and its Applications }}.
\newblock Wiley-Interscience New York, 1971.

\bibitem{McCulloch_Pitts}
W.~S. McCulloch and W.~Pitts.
\newblock {\em A Logical Calculus of the Ideas Immanent in Nervous Activity}.
\newblock MIT Press, Cambridge, MA, USA, 1988.

\bibitem{Samuel_2010_ICM}
S.~Leshner, B.~Krzysztof, and S.~Vrudhula.
\newblock Design of a robust, high performance standard cell threshold logic
  family for deep sub-micron technology.
\newblock In {\em Proceedings of the IEEE International Conference on
  Microelectronics}, Cairo, Egypt, Dec. 19-22 2010.

\bibitem{Neutzling_ICCAD_2015}
A.~Neutzling, J.~M. Matos, A.~I. Reis, R.~P. Ribas, and A.~Mishchenko.
\newblock {Threshold logic synthesis based on cut pruning}.
\newblock In {\em IEEE/ACM ICCAD}, Nov 2015.

\bibitem{Kulkarni_TVLSI_2016}
N.~Kulkarni, J.~Yang, J.~S. Seo, and S.~Vrudhula.
\newblock {Reducing Power, Leakage, and Area of Standard-Cell ASICs Using
  Threshold Logic Flip-Flops}.
\newblock {\em IEEE TVLSI}, 24(9), Sept 2016.

\bibitem{ANNAMPEDU201384}
Viswanath Annampedu and Meghanad~D. Wagh.
\newblock {Decomposition of threshold functions into bounded fan-in threshold
  functions}.
\newblock {\em Information and Computation}, 227:84--101, 2013.

\bibitem{simons2019review}
Taylor Simons and Dah-Jye Lee.
\newblock A review of binarized neural networks.
\newblock {\em Electronics}, 8(6):661, 2019.

\bibitem{2016_Eyeriss}
Y.~{Chen}, J.~{Emer}, and V.~{Sze}.
\newblock Eyeriss: A spatial architecture for energy-efficient dataflow for
  convolutional neural networks.
\newblock In {\em 2016 ACM/IEEE ISCA}, pages 367--379, 2016.

\bibitem{xnornet}
M.~Rastegari, V.~Ordonez, J.~Redmon, and A.~Farhadi.
\newblock Xnor-net: Imagenet classification using binary convolutional neural
  networks.
\newblock volume 9908, pages 525--542, 10 2016.

\end{thebibliography}
		
	\end{spacing}
	
\end{document}